\documentstyle[prl,aps]{revtex}

\begin{document}
\draft
\twocolumn

\title{Mixed-state entanglement and distillation: is there  a ``bound''
entanglement in nature?}

\author{Micha\l{} Horodecki\cite{poczta1}}

\address{Institute of Theoretical Physics and Astrophysics\\
University of Gda\'nsk, 80--952 Gda\'nsk, Poland}

\author{Pawe\l{} Horodecki\cite{poczta2}}

\address{Faculty of Applied Physics and Mathematics\\
Technical University of Gda\'nsk, 80--952 Gda\'nsk, Poland}

\author{Ryszard Horodecki\cite{poczta}}

\address{Institute of Theoretical Physics and Astrophysics\\
University of Gda\'nsk, 80--952 Gda\'nsk, Poland}

\maketitle

\begin{abstract}
It is  shown that if a mixed state can be distilled to the singlet form,
it must violate partial transposition criterion
[A. Peres, Phys. Rev. Lett. {\bf 76}, 1413 (1996)]. It implies that there
are two {\it qualitatively} different types of entanglement: 
`` free'' entanglement which is distillable, and ``bound'' entanglement
which cannot be brought to
the singlet form useful for quantum communication purposes.
Possible physical meaning of the result is discussed.
\end{abstract}
\pacs{Pacs Numbers: 03.65.Bz}

Since the famous Einstein, Podolsky and Rosen  \cite{EPR} and Schr\"odinger
\cite{Sch}
papers quantum entanglement still remains one of the most striking
implications of quantum formalism. In recent years,
a great effort was made to understand a role of entanglement in nature and
fundamental applications were found in the field of quantum information
theory \cite{Ekert,geste,Bennett_tel,huge}.
The most familiar example of pure entangled state is the singlet
state \cite{Bohm} of two spin-$1\over2$ particles
\begin{equation}
\Psi_-={1\over \sqrt2} (|\uparrow\downarrow\rangle-|\downarrow\uparrow\rangle),
\end{equation}
which cannot be reduced to direct product by any
transformation of the bases pertaining to each one of the particles.

In practice, due to decoherence effects, we usually deal with mixed states
\cite{states}.
A mixed state of quantum system consisting of two subsystems
is supposed to represent entanglement if it is
inseparable \cite{Werner} i.e. cannot be written in the form
\begin{equation}
\varrho=\sum_ip_i\varrho^A_i\otimes \varrho^B_i,\quad p_i\geq0,
\quad \sum_ip_i=1.
\label{def}
\end{equation}
were $\varrho^A_i$ and $\varrho^B_i$ are states for the two subsystems.
However, to use the entanglement for quantum information processing,
we must have it in pure singlet form.  The procedure of converting
mixed state entanglement to the singlet form is called distillation
\cite{Bennett_pur}. It amounts to extraction of pairs \cite{trzy}
of particles in singlet state from an ensemble described by  some
mixed state by means of local quantum operations and classical communication
\cite{Bennett_pur}. 

The process can be described as follows :
the two observers, Alice and Bob, each have
N quantum systems coming from entangled pairs prepared in a given state
$\rho$. Each one can perform local operations with her/his N particles,
and exchange classical information with the other one. The question is
whether they can in this way obtain a pair of entangled qubits (the rest
of the quantum systems being discarded). They need not succeed every
time, but at least they know when they have been successful.
If they managed to do this, one says that they have {\it distilled}
some amount of pure entanglement from the state $\varrho$. 
Subsequently, the distilled singlet pairs can be used e.g. for 
reliable transmission of quantum information via teleportation 
\cite{Bennett_tel}.

Recently, it has been  shown
\cite{pur} that {\it any} inseparable two-qubit state \cite{qubit}
represents the
entanglement which, however small, can be distilled to a singlet form.
The result was obtained  by use of  the  necessary \cite{Peres} and
sufficient \cite{sep}
condition of separability for  two-qubit states, local
filtering \cite{Gisin,conc} and Bennett {\it et al.} distillation
protocol \cite{Bennett_pur}i. 

In this context it seems very natural to make the following conjecture:

{\it Conjecture -} Any inseparable state can be distilled to the singlet form.

Surprisingly enough, this conjecture is wrong.
In the present Letter we will show that there  are inseparable
states that {\it cannot} be distilled. More specifically, we first show that
any state which can be distilled must violate Peres separability criterion
 \cite{Peres}. Then the result follows from
the fact \cite{Pawel} that there are inseparable states that  satisfy
the criterion. It shows that there are two  qualitatively
different types of entanglement. The first,  
``free'' entanglement, can be distilled to the singlet form.  
The second type of entanglement is not distillable and 
is considered here in analogy with thermodynamics as a ``bound'' 
entanglement which cannot be used to perform a useful ``informational work'' 
like reliable transmission of quantum data via teleportation.

Now, let us first shortly describe the Peres criterion.
A state $\varrho$ satisfies the criterion,
if all eigenvalues of its partial transposition $\varrho^{T_B}$
are nonnegative (i.e. if $\varrho^{T_B}$ is a positive operator).
Here the partial transposition $\varrho^{T_B}$
associated with an  arbitrary product orthonormal $e_i \otimes f_j$
basis is defined by the matrix elements in this basis:
\begin{equation}
\varrho^{T_B}_{m\mu,n\nu}\equiv
\langle e_m\otimes f_\mu| \varrho^{T_B}| e_n \otimes f_\nu\rangle=
\varrho_{m\nu,n\mu}.
\end{equation}
Clearly, the matrix $\varrho^{T_B}$ depends on
the basis, but its eigenvalues do not.
Thus given a state, one can check whether it violates the criterion
performing the partial transposition in an arbitrary product basis.
In particular, it implies that
$\varrho$ violates the criterion if and only if any N-fold tensor
product $\varrho^{\otimes N}=\underbrace{\varrho\otimes
\ldots \otimes\varrho}_N$  does \cite{Peres}.

Peres showed that the  criterion must be satisfied by
any separable state \cite{Peres}. It has been also shown \cite{sep}
that for two-qubit (and qubit-trit)
states the criterion is also {\it sufficient} condition for
separability. This does {\it not} hold for higher dimensions. The explicit
examples of inseparable mixtures satisfying criterion were constructed
\cite{Pawel}.

Now we are in position to present the main result of this Letter.
Suppose Alice and Bob have a large number $N$ of pairs each in
a state $\varrho$ acting on the Hilbert space ${\cal H}
={\cal H}_A\otimes {\cal H}_B$. Then the joint state of $N$ pairs
is given by $\varrho^{\otimes N}$.
Suppose now that the state $\varrho$ is distillable.   This means, that Alice
and Bob are able to obtain pure singlet
two-qubit pairs for $N$ tending to infinity. This however implies, that
for some finite $N$, they are able to obtain an inseparable two-qubit state
$\tilde\varrho_{2q}$.
The most general operation producing a two-qubit pair they can perform
over the initial amount of $N$ pairs can be written in the following form
\cite{Knight}
\begin{equation}
\tilde\varrho_{2q}=
{1\over M}\sum_i A_i\otimes B_i \varrho^{\otimes N} A^\dagger_i
\otimes B^\dagger_i,
\label{superoperator}
\end{equation}
where $M={\rm Tr}\sum_i A_i\otimes B_i \varrho^{\otimes N} A^\dagger_i
\otimes B^\dagger_i$ is the normalization factor and
$A_i$ and $B_i$ map the large Hilbert spaces ${\cal H}_{A,B}^{\otimes N}$
into $C^2$.
For convenience, we will use unnormalized states, as the property
of separability  as well as satisfying the Peres criterion
do not depend on the positive factor.
Then, for unnormalized states, we omit the condition $\sum_ip_i=1$ in the
definition of separability (\ref{def}).
Consequently  let
\begin{equation}
\varrho_{2q}=\sum_i A_i\otimes B_i \varrho^{\otimes N} A^\dagger_i
\otimes B^\dagger_i
\end{equation}
and
\begin{equation}
\varrho_i=A_i\otimes B_i \varrho^{\otimes N} A^\dagger_i
\otimes B^\dagger_i.
\end{equation}
Since $\varrho_{2q}$ is inseparable  then at least for some $i=i_0$ the
state $\varrho_{i_0}$ must be inseparable. Indeed, by summing separable states
we cannot get inseparable one.

Note that
the operators $A_{i_0}$ and $B_{i_0}$ act into two-dimensional space $C^2$,
hence they can be written in the  form
\begin{equation}
A_{i_0}=|0\rangle\langle\psi_A| +|1\rangle\langle\phi_A|, \quad
B_{i_0}=|0\rangle\langle\psi_B| +|1\rangle\langle\phi_B|,
\end{equation}
where $|1\rangle$ and $|0\rangle$ constitute orthonormal basis in $C^2$ and
$\psi_A, \phi_A\in {\cal H}_A^{\otimes N}$,
$\psi_B, \phi_B\in {\cal H}_B^{\otimes N}$ are arbitrary
(possibly unnormalized) vectors. Let us now
consider two-dimensional projectors $P_A$ and $P_B$
which project onto the spaces spanned by $\psi_A, \phi_A$ and
$\psi_B, \phi_B$ respectively. Then we have
\begin{equation}
\varrho_{i_0}=A_{i_0}\otimes B_{i_0} \left(P_A\otimes P_B\varrho^{\otimes N}
 P_A\otimes P_B \right)A^\dagger_{i_0} \otimes B^\dagger_{i_0}.
\end{equation}
Now, since a  product action cannot convert separable state into 
inseparable one, we obtain that also the state
\begin{equation}
\varrho'=P_A\otimes P_B\varrho^{\otimes N}
P_A\otimes P_B
\label{roprim}
\end{equation} is inseparable.
%%%%%%%%%%%%%%%%%%%%%%%%%%%%%%%
Let us write this state in basis
$| f_i \rangle \otimes | g_k \rangle, i= 1, 2 , ...,dim {\cal H}_A^{\otimes N},
k= 1, 2 , ...,dim {\cal H}_B^{\otimes N} $ with four vectors
$| f_1 \rangle, | f_2 \rangle $ ($| g_1 \rangle, | g_2 \rangle $)
spanning the subspaces defined by projectors $P_A$, $P_B$.
The only nonzero matrix elements  are due to products of those vectors
and they define a $4 \times 4$ matrix $M_{2q}$ which can be thought
as two-qubit state. The operation of partial transposition on $\varrho'$
affects only those elements (as the remaining ones are equal to zero).
If $M_{2q}$ were positive after partial transposition, then,
due to the sufficiency
of the partial transposition test for two-qubit case \cite{sep},
$M_{2q}$ would represent a separable two-qubit state.
Hence, if embedded into
the whole space ${\cal H}^{\otimes N}$, it would still remain separable.
Consequently,
the state $\varrho'$ would be separable, which is the contradiction.
Thus partial transposition of $M_{2q}$ must be negative. Now, since
$M_{2q}$ is formed by all nonzero elements of $\varrho'$, then
we obtain that also the state $\varrho'$ must violate the Peres
criterion,
i. e. $\varrho'^{T_B}$ must have a negative eigenvalue.
Now let $\psi$ be the eigenvector corresponding to the eigenvalue.
As the vector belongs to the subspace ${\cal H}_{2q}$
it follows that the matrix elements
$\langle\psi|\varrho'^{T_B}|\psi\rangle$ and
$\langle\psi|({\varrho^{\otimes N}})^{T_B}|\psi\rangle$ are equal. Hence
we obtain
\begin{equation}
\langle\psi|({ \varrho^{\otimes N}})^{T_B}
|\psi\rangle<0.
\end{equation}
Thus the state $\varrho^{\otimes N}$ violates the partial
transposition criterion. However,
as it was mentioned, this implies that also $\varrho$ does. 
All the above consideration can be formally summarised as follows. 
If the output state of this action appears to have negative 
partial transposition, then
the basic component $\varrho $ of input state $\varrho ^{\otimes N}$
must have had also negative partial transposition . 
This means nothing but that any
action of type (\ref{superoperator}) on $\varrho$ (including collecting 
N pairs) preserves positivity of partial transposition. 
This result can be generalized \cite{Asher}:
any action of  the form   
${1 \over M} 
\sum_i A_i\otimes B_i \varrho^{\otimes N} A^\dagger_i \otimes B^\dagger_i$
producing an arbitrary two-component system (not necessarily
$2 \times 2$ one) preserves positivity of partial transposition.  

Thus we showed that if a state $\varrho$ is distillable, it must
violate the Peres separability criterion.
It is an important result as it implies that there are inseparable states
which {\it cannot} be  distilled! Indeed, quite recently one of us \cite{Pawel}
constructed inseparable states which do not violate the criterion.
Some of those peculiar states are density matrices for two spin-1
particles (the two-trit case).
Using the standard basis for this case
( $| 1 \rangle | 1 \rangle , | 1 \rangle | 2 \rangle ,
| 1 \rangle | 3 \rangle , | 2 \rangle | 1 \rangle , | 2 \rangle | 2 \rangle $,
and so on ... ) those matrices can be written in the form:
\begin{eqnarray}
\varrho_a={1 \over 8a + 1}
\left[ \begin{array}{ccccccccc}
	  a &0&0&0&a&0&0&0& a	\\
	   0&a&0&0&0&0&0&0&0	 \\
	   0&0&a&0&0&0&0&0&0	 \\
	   0&0&0&a&0&0&0&0&0	 \\
	  a &0&0&0&a&0&0&0& a	  \\
	   0&0&0&0&0&a&0&0&0	 \\
	   0&0&0&0&0&0&{1+a \over 2}&0&{\sqrt{1-a^2} \over 2}\\
	   0&0&0&0&0&0&0&a&0	 \\
	  a &0&0&0&a&0&{\sqrt{1-a^2} \over 2}&0&{1+a \over 2}\\
       \end{array}
      \right ], \ \ \
\label{tran}
\end{eqnarray}
with $0 < a < 1$.
It has been shown \cite{Pawel} by means of independent separability criterion
that those states are {\it inseparable} despite they have {\it positive}
partial transposition. However, as we have shown above, that the density
matrices with positive partial transposition {\it cannot} be distilled to
the singlet form. Consequently, any  state of the form (\ref{tran}) cannot
be distilled.

It is remarkable, that the question
whether a state is distillable or not has been reduced to the one whether
there is a two-qubit entanglement in a collection of $N$ pairs for some $N$.
Thus the latter condition is {\it the necessary and sufficient condition}
for any given state to be distilled.  
Indeed, as shown above, if a state $\varrho$ is distillable then there
exist two-dimensional projections $P_A$ and $P_B$ so that the state $\varrho'$
given by eq. (\ref{roprim}) is inseparable. Conversely, if the
latter condition is satisfied
then $\varrho$ can be distilled by projecting $\varrho^{\otimes N}$
locally by means of $P_A$ and $P_B$ and then applying the protocol proposed
in \cite{pur} which is able to distill any two-qubit inseparable state.
There is an open question, whether the condition implies
satisfying Peres criterion. Then the latter would acquire the physical sense:
it would be equivalent to distillability.

Let us now discuss shortly possible physical meaning of our result.
As a matter of fact, we have revealed a kind of entanglement which
cannot be used for sending  reliably quantum information via teleportation.
Using an analogy with thermodynamics \cite{ther}, we can consider
entanglement as a counterpart of energy, and sending of quantum information
as a kind of ``informational work''. Consequently we can consider
``free entanglement'' ($E_{free}$) which can be distilled,
and ``bound entanglement''
($E_{bound}$). In particular, the free entanglement is naturally identified
with distillable entanglement $D$ as the latter says us how much qubits can
we reliably teleport via the mixed state.  This kind of entanglement  can be
always converted via distillation protocol to the  ``active'' singlet form.

To complete the analogy, one could consider the asymptotic
number of singlets which are needed to produce a given mixed state
as internal entanglement $E_{int}$ (the counterpart of internal energy)
\cite{formation}.
Then the bound entanglement can be quantitatively defined by the
following equation
\begin{equation}
E_{int}=E_{free}+E_{bound}.
\end{equation}
In particular, for pure states we have $E_{int}=E_{free}$ and $E_{bound}=0$.
Indeed, pure states can be converted in a ``lossless''  way into
active singlet form \cite{conc}.
In the present letter we showed that there exist {\it inseparable}
states having reciprocal properties. Namely for the states of
type (\ref{tran}) we have $E_{int}=E_{bound}$ and $E_{free}=0$.

Now the  question arises:  is it that $E_{int}=E_{bound}=0$ or $\not=0$?
Both cases are curious. In the first case, we would have {\it inseparable}
states which can be produced from asymptotically {\it zero} number of singlet
pairs. This would imply, in turn, that entanglement of formation is {\it not}
additive state function \cite{Wooters}, as by the very definition
it does not vanish for any inseparable states.
In the second case, we would have
curious states which absorb entanglement in an irreversible way.
To produce  such states, one needs some amount of entanglement. But once
the states were produced, there is no way to recover any, however little,
piece of the initial entanglement. The latter is {\it entirely} lost.

A natural problem which arises  in the context of the presented result is:
what is  the physical reason  for which the partial transposition is connected
with distillability? Our conjecture is that it is {\it time} which links
intimately the two things.
Indeed, transposition can be interpreted as the operation of
time-reversal \cite{Busch}.
Also in the context of distillation, there appeared the problem of time.
Namely, distillation is inherently connected with the quantum error
correction for
quantum noisy channel supplemented by two-way classical channel \cite{huge}.
The quantum capacity of such channels can be strictly larger than
without the classical channel. However, the price we
must pay is that the error correction with two-way classical communication
cannot be used to store the quantum information in noisy environment
\cite{huge} because one cannot send signal backward in time.
%Needless to say the connection between the distillation, partial transposition
%and time reversal requires a deeper investigation.
Needless to say deeper investigation of the connection among
the distillation, partial transposition and time reversal seems
to be more than desirable.

Finally, it is perhaps worth to mention about the circle described by
story of the nonlocality of mixed states,
beginning with the work of Werner \cite{Werner}. The latter suggested
that there are curious inseparable states which do not exhibit nonlocal
correlations. Then Popescu \cite{Popescu_hid} showed that there is a subtle
kind of pure
quantum correlations which is exhibited by Werner mixtures. The distillability
of all two-qubit states \cite{pur} proved that all they are also nonlocal.
One could suspect that the story will end by showing that all
inseparable states can be distilled, hence they are nonlocal. Here
we showed that it is not true. So, one is
now faced with the problem similar to the initial one i.e. are
the inseparable states with positive partial transposition nonlocal?
Now, in view of the above result it follows that the problem certainly
cannot be solved by means of distillation concept.

We would like to thank Asher Peres for helpful comments 
and discussion.


\begin{references}
\bibitem{states}
In this Letter we identify  states with their density matrices.
\bibitem{qubit}
In analogy to classical bits, a two-level (spin-$1\over2$) quantum system
is usually referred to as qubit. Similarly, a three level (spin-$1$)
system is called trit.
\bibitem{formation}
One of the important problems is finding whether the so defined internal
entanglement is equal to the entanglement of formation \cite{Wooters}.
The latter is defined  \cite{huge} by
$E(\varrho)=\inf\sum_i p_iS(\varrho_i)$,
where $S(\varrho)=-{\rm Tr} \varrho\log_2\varrho$. The infimum is taken
over all ensembles of pure states $\psi_i$ satisfying
$\varrho_i={\rm Tr_B} |\psi_i\rangle\langle\psi_i|$, and
$\varrho=\sum_ip_i|\psi_i\rangle\langle\psi_i|$.
It can be seen that $E_{int}(\varrho)=\lim_N {1\over N} 
E(\varrho^{\otimes N})$
(we could say that $E_{int}$ is ``additification'' of $E$).
\end{references}
\end{document}